\documentclass[aps,prd,eprint,numerical,showpacs,showkeys,10pt]{revtex4-1}

\usepackage{amsmath}
\usepackage{amssymb}
\usepackage{amsfonts}
\usepackage{graphicx}
\usepackage{dcolumn}
\usepackage{bm}
\usepackage{hyperref}
\hypersetup{colorlinks=false}

\begin{document}

\title[Lovelock black hole: strong homotopy retract and quasistationary levels]{Some aspects of the five-dimensional Lovelock black hole spacetime: strong homotopy retract, perihelion precession and quasistationary levels}

\date{\today}

\author{M. Abu-Saleem}
\email{mohammedabusaleem2005@yahoo.com and m\_abusaleem@bau.edu.jo}
\affiliation{Mathematics Department, Faculty of Science, Al-Balqa Applied University, 19117 Salt, Jordan}
\author{H. S. Vieira}
\email{horacio.santana.vieira@hotmail.com and horacio.santana-vieira@tat.uni-tuebingen.de}
\affiliation{Theoretical Astrophysics, Institute for Astronomy and Astrophysics, University of T\"{u}bingen, 72076 T\"{u}bingen, Germany}

\begin{abstract}
In this work we explore some mathematical physics aspects of the spherically symmetric Lovelock black hole in high dimensions. Intended for this aim, we thoroughly consider the metric corresponding to the five-dimensional Lovelock black hole spacetime. We construct the strong retractions by the geodesic equations on the background under consideration. As a result, from the topological point of view, we construct the theory of strong homotopy retract, which will allow us, in principle, to better understand some of its suitable applications on astrophysics and cosmology, in particular, in the analysis of the spacetime singularities. We find the solutions of the equation of motion for both radial and angular coordinates, and then we describe the outer (``exterior'') and lower (``interior'') apparent horizons. Indeed, the outer apparent horizon is the last surface from which the light waves could still escape from the black hole. Thus, it is meaningful to analyze some physical phenomena related to quantum particles propagating outside the exterior apparent horizon, in particular, we discuss the quasistationary levels of scalar fields and their radial wave functions, which are given in terms of the general Heun functions. We also calculate the perihelion precession in this background.
\end{abstract}



\preprint{Preprint submitted to Annals of Physics}
%\preprint{AIP/123-QED}

\maketitle

%\begin{quotation}
%...
%\end{quotation}

%
%%%%%%%%%%%%%%%%%%%%%%%%%%%%%%%%%%%%%%%%%%%%%%%%%%%%%%%%%%%%%%%%%%%%%%%%%%%%%%%%%%%%%%%%%%%%%% Introduction
%
\section{Introducing the five-dimensional Lovelock black hole spacetime}
From a mathematical point of view, general relativity is a theory of geometries. On the other hand, our daily physics is positioned in the scenario of a specific spacetime, namely, the special relativity. The general theory of relativity relates the existence of a gravitational field with the nearby spacetime curvature. For susceptible fields, such as show up on the earth's surfaces, the curvature is extraordinarily insignificant, and for all aims, our nearby physics takes place in a flat (Minkowski) spacetime. The special theory of relativity is the particular case of the general theory of relativity to the Minkowski spacetime. The spacetime curvature is certainly illustrated by using the so-called line element. The line element for a specific distribution of matter and/or energy is obtained by solving the Einstein's field equations. These well-known equations are a set of highly nonlinear and coupled partial differential equations \cite{Wild} (in addition, for recent critical overviews of general relativity, see Refs.~\cite{a,b,c,d} and references therein).

On the other hand, the Lovelock theory of gravity is an actual broadening of Einstein's relativity to higher dimensions, and it plays a very important role in theoretical (mathematical) physics, as it provides a large category of models, as for example, the topological black hole solutions, which characterize some black branes of the theory, as well as the vacuum thin-shell wormhole-like geometries that join two distinct asymptotically de Sitter spacetimes. Therefore, the Lovelock theory of gravity characterizes a very fascinating scenario, where is possible, in principle, to find out about how to study the physical results of a corrected theory of gravity at short distances, as the result of existing higher-order curvature conditions in the action \cite{Garra}.

It is well known that the general theory of relativity and the quantum mechanics are incompatible in their current form. However, after Hawking \cite{CommMathPhys.43.199} found that black holes can emit, scatter and absorb radiation, as well as that the evaporation rate is proportional to the total absorption cross-section, a lot of works have been published in this line of research, namely, the interaction between quantum fields and black holes (for a review, see Chen \textit{et al.} \cite{Chen} and references therein). This line of research constitutes one of several attempts that try to construct a complete theory of quantum gravity.

The 1915 Einstein's general theory of relativity can be generalized to higher dimensions by keeping almost its characteristics. As a result, we get the so-called Lovelock theory of gravity, which provides second-order field equations in arbitrary dimensions. The equation of motion in a Lovelock theory of gravity can be seen as the low energy limit of a heterotic superstring theory \cite{Ghos}. In this approach, the line element describing a D-dimensional Schwarzschild black hole spacetime in the string cloud model is given by
\begin{equation}
ds^{2}=-f(r)\ dt^{2}+[f(r)]^{-1}\ dr^{2}+r^{2}\ d\Omega_{D-2}^{2},
\label{e(1)}
\end{equation}
with
\begin{equation}
f(r)=1-\frac{2M}{(D-3)r^{D-3}}-\frac{2a}{(D-2)r^{D-4}},
\label{e(2)}
\end{equation}
and
\begin{equation}
d\Omega_{D-2}^{2}=\gamma_{ij}dx^{i}dx^{j},
\label{e(3)}
\end{equation}
where $D$ is the dimension of the spacetime, $M$ is the spherical mass centered at the origin of such a coordinate system, and $a$ is the constant associated with the stress tensor of the string cloud model. The angular line element $d\Omega_{D-2}^{2}$ is associated with a $(D-2)$-dimensional hypersurface with constant curvature, which correspond to the hyperbolic, flat or spherical spaces for $k=-1$, $0$ or $+1$, respectively. In this work, we will deal with a five-dimensional spherically symmetric spacetime \cite{Habib,Letel,Lovel,Meng,Vieira}, so that
\begin{equation}
d\Omega_{3}^{2}=d\phi^{2}+\sin^{2}\phi\ (d\theta^{2}+\sin^{2}\theta\ d\varphi^{2}).
\label{e(4)}
\end{equation}
This angular line element describes a three-sphere, where $0 \leq \phi \leq \pi$, $0 \leq \theta \leq \pi$ and $0 \leq \varphi \leq 2\pi$. Thus, the metric of a five-dimensional Lovelock black hole spacetime can be written as \cite{Vieira}
\begin{equation}
ds^{2}=-f(r)\ dt^{2}+[f(r)]^{-1}\ dr^{2}+r^{2}\ [d\phi^{2}+\sin^{2}\phi\ (d\theta^{2}+\sin^{2}\theta\ d\varphi^{2})],
\label{e(5)}
\end{equation}
where
\begin{equation}
f(r)=1-\frac{M}{r^{2}}-\frac{2a}{3r}.
\label{e(6)}
\end{equation}
The black hole horizons are the solutions of the surface equation $f(r)=0$. They are given by
\begin{equation}
r_{\mbox{\tiny{AH}}} = \frac{1}{3}(a \pm \sqrt{a^{2}+9M}),
\label{eq:horizons_Lovelock_D5}
\end{equation}
where the subscript $\mbox{\tiny{AH}}$ means ``Apparent Horizon''. In fact, the background under consideration has apparent horizons (not event horizons), which are the outermost marginally trapped surface for the outgoing photons (for a detailed review about the radiating black hole horizons, see \cite{Vieira} and references therein). However, for simplicity and convenience, without loss of generality, we can also refer to these solutions as the exterior, $r_{+}$, and interior, $r_{-}$, ``event'' horizons.

Thus, since the physics of black holes is mainly related to the wave phenomena occurring outside the exterior apparent horizon, it is meaningful to study the existence of such a singularity in the background under consideration. Afterward, we will discuss a very special wave phenomena, namely, the quasistationary levels of scalar particles.

This paper has two broad aims. First, to present some results on the topological aspects of the five-dimensional Lovelock black hole spacetime, and then use it to proof the existence of the apparent horizons, which is crucial to the studies of wave phenomena on black holes. Second, to discuss the interaction between quantum scalar particles and the background under consideration, and then obtains the quasistationary levels, which are related to the boundary conditions imposed on the radial wave solution. In fact, these are two very interesting features in a mathematical physics point of view.

The remainder of this paper is organized as follows. In Section \ref{Geometric}, we explore some topological aspects of both strong and induced strong homotopy retractions. In Section \ref{Perihelion}, we calculate the (anomalous) perihelion precession which comes out from the model considered. In Section \ref{Quasistationary}, we obtain the radial wave solution and discuss the scalar resonant frequencies related to the quasistationary levels. Finally, in Section \ref{Conclusions}, we summarize our results.
%
%We will also discuss the information near the black hole apparent horizon, which is conserved under a suitable transformation of the basic creation and annihilation operators.
%
%
%%%%%%%%%%%%%%%%%%%%%%%%%%%%%%%%%%%%%%%%%%%%%%%%%%%%%%%%%%%%%%%%%%%%%%%%%%%%%%%%%%%%%%%%%%%%%% Geometric and homotopy retracts
%
\section{Geometric and homotopy retracts}\label{Geometric}
It is known that geometric topology is the study of manifolds and maps between them, particularly embeddings of one manifold into another. On the other hand, a retraction is said to be a continuous mapping from a topological space into a subspace by preserving the position of all points in that subspace; in this case, the subspace is called a retract of the original space. A deformation retraction is a mapping that captures the idea of continuously shrinking a space into a subspace. In particular, an elastic homotopy is the studies of shape up until to the equivalent deformities that may be stretching, shrinking, bending, or twisting but not cut or glue \cite{Fox}.

In addition, the subspace of a topological space is termed as a homotopy retract if the identity map, from the whole space to itself, is homotopic to the retraction onto that subspace. Thus, let $\mathcal{D}$ be a subspace of $\mathcal{X}$. Then, $\mathcal{D}$ is called a retract of $\mathcal{X}$ if there is a continuous map $y:\mathcal{X} \rightarrow \mathcal{X}$ (called a retraction) such that, for all $x \in \mathcal{X}$ and all $d \in \mathcal{D}$, it is true $y(x) \in \mathcal{D}$ and $y(d)=d$ \cite{Massey}.

Therefore, the theory of retracts can be described as the principle of some kind of topological properties, associated with a class of functions. These maps are much more general than homeomorphism, as well as much more special than arbitrary continuous maps. This theory also deals with many other topological properties, where many of them have a clear-cut geometrical aspect.

In what follows, we define some fascinating new requirements in topology, specifically, the strong retraction and the strong homotopy retract. Then, as a result, we will apply this theory to the metric of a five-dimensional Lovelock black hole spacetime.
%
%%%%%%%%%%%%%%%%%%%%%%%%%%%%%%%%%%%%%%%%%%%%%%%%%%%%%%%%%%%%%%%%%%%%%%%%%%%%%%%%%%%%%%%%%%%%%% Some definitions
%
\subsection{Some definitions}
First of all, we will establish some definitions related to the strong and strong homotopy retracts, as follows.

%\begin{Definition}
\noindent \textbf{Definition 1.} Let $U$ be an $n$-dimensional manifold. Let $U_{0} \subseteq U$ be given the subspace topology. A strong retraction is a continuous map $\varpi:U \rightarrow U_{0}$ such that\\
(i) $U$ is open,\\
(ii) $\varpi (x) = x$ ($\forall\ x \in U_{0}$),\\
(iii) $\varpi (U) = U_{0}$,\\
(iv) $\varpi (U)$ is a manifold with constant curvature.
%\label{def1}
%\end{Definition}

%\begin{Definition}
\noindent \textbf{Definition 2.} A subset $U_{0}$ of a manifold $U$ is said to be a strong homotopy retracts if there exists a strong retraction $\varpi:U \rightarrow U_{0}$ and a homotopy $\mathfrak{H}:U \times [0,1] \rightarrow U$ such that\\
(i) $\mathfrak{H} (x,0) = x$ ($\forall\ x \in U$),\\
(ii) $\mathfrak{H} (x,1) = \varpi (x)$ ($\forall\ x \in U$),\\
(iii) $\mathfrak{H} (x_{0},\varpi) = x_{0}$ ($\forall\ x_{0} \in U_{0}$ and $\forall\ \varpi \in [0,1]$).
%\label{def2}
%\end{Definition}

In what follows, we will obtain some topological, kinematical and dynamical quantities of the five-dimensional Lovelock black hole spacetime.
%
%%%%%%%%%%%%%%%%%%%%%%%%%%%%%%%%%%%%%%%%%%%%%%%%%%%%%%%%%%%%%%%%%%%%%%%%%%%%%%%%%%%%%%%%%%%%%% Some parameters of the five-dimensional Lovelock black hole
%
\subsection{Some parameters of the five-dimensional Lovelock black hole spacetime}
In order to determine three styles of strong retractions the five-dimensional Lovelock black hole spacetime, let us write some important parameters of this background and then we will use the Lagrangian formalism to get a set of geodesic equations.

The most general metric of a five-dimensional black hole solution can be written as
\begin{equation}
ds^{2}=-du_{1}^{2}+du_{2}^{2}+du_{3}^{2}+du_{4}^{2}+du_{5}^{2}.
\label{eq:general_metric}
\end{equation}
Now, let $\mathrm{V}^{5}$ be the five-dimensional Lovelock black hole spacetime. The parameters $u_{i}$ ($i=1,\ldots,5$) of $\mathrm{V}^{5}$ can be obtained by comparing Eqs.~(\ref{e(5)}) and (\ref{eq:general_metric}). This will lead to a system of differential equations that must be solved to introduce the following parameters
\begin{eqnarray}
u_{1} & = & \pm \sqrt{f(r)}t+b_{1},\nonumber\\
u_{2} & = & \pm \biggl\{\frac{r}{\sqrt{3}}\sqrt{\frac{2a-6M+3r}{r}}+\biggl(M-\frac{a}{3}\biggr)\log\biggl(r\sqrt{\frac{6a-18M+9r}{r}}+3\biggr)+a-3M\biggr\}+b_{2},\nonumber\\
u_{3} & = & \pm r\phi+b_{3},\nonumber\\
u_{4} & = & \pm (r\sin\phi)\theta+b_{4},\nonumber\\
u_{5} & = & \pm (r\sin\phi\sin\theta)\varphi+b_{5},
\label{e(8)}
\end{eqnarray}
where $b_{i}$ ($i=1,\ldots,5$) are constants (of integration) to be determined.

Now, for a given metric $g_{\mu\nu}$, the geodesic equations can be obtained from the Lagrangian, $\mathcal{L}$, and the Euler-Lagrange equation, which are given by
\begin{equation}
\mathcal{L}=\frac{1}{2}g_{\mu\nu}\dot{x}^{\mu}\dot{x}^{\nu}
\label{e(9)}
\end{equation}
and
\begin{equation}
\frac{d}{ds}\biggl(\frac{\partial\mathcal{L}}{\partial \dot{x}^{\mu}}\biggr)-\frac{\partial\mathcal{L}}{\partial x^{\mu}}=0,
\label{e(10)}
\end{equation}
where the dot ``$\ \dot{}\ $'' represents the derivative related to the proper length $s$ and $\mu,\nu=1,\ldots,5$. Thus, the Lagrangian for $\mathrm{V}^{5}$ reads
\begin{equation}
\mathcal{L}=\frac{1}{2}\biggl\{-f(r)\ \dot{t}^{2}+\frac{1}{f(r)}\ \dot{r}^{2}+r^{2}[\dot{\phi^{2}}+\sin^{2}\phi\ (\dot{\theta}^{2}+\sin^{2}\theta\ \dot{\varphi}^{2})]\biggr\}.
\label{e(11)}
\end{equation}
On the other hand, from the Euler-Lagrange equation, we get the following equations for the temporal $t$ and radial $r$ coordinates of $\mathrm{V}^{5}$:
\begin{equation}
f(r)\ \dot{t}=c_{1},
\label{e(12)}
\end{equation}
and
\begin{equation}
\frac{d}{ds}\biggl[\frac{\dot{r}}{f(r)}\biggr]+\biggl(\frac{M}{r^{3}}+\frac{a}{3r^{2}}\biggr)\dot{t}^{2}+\frac{\dot{r}^{2}}{[f(r)]^{2}}\biggl(\frac{M}{r^{3}}+\frac{a}{3r^{2}}\biggr)+r[\dot{\phi}^{2}+\sin^{2}\phi(\dot{\theta}^{2}+\sin^{2}\theta\ \dot{\varphi}^{2})]=0.
\label{e(13)}
\end{equation}
Furthermore, for the angular coordinates $\theta$, $\phi$ and $\varphi$, we obtain
\begin{equation}
\frac{d}{ds}(r^{2}\sin^{2}\phi\ \dot{\theta})-\frac{r^{2}\sin^{2}\phi\ \sin(2\theta)\ \dot{\varphi}^{2}}{2}=0,
\label{e(14a)}
\end{equation}
\begin{equation}
\frac{d}{ds}(r^{2}\dot{\phi})-\frac{\sin(2\phi)\ \dot{\theta}^{2}}{2}=0,
\label{e(15a)}
\end{equation}
\begin{equation}
r^{2}\sin^{2}\phi\ \sin^{2}\theta\ \dot{\varphi}=c_{2},
\label{e(16a)}
\end{equation}
where $c_{1}$ and $c_{2}$ are constants (of integration) to be determined.

From Eq.~(\ref{e(12)}), in the special case when $c_{1}=0$, we can obtain the following solution to the time $t=c_{3}$, where $c_{3}$ is also a constant to be determined. In addition, for the particular case when $f(r)=0$, it yields the surfaces $r_{\pm}=(a \pm \sqrt{a^{2}+9M})/3$, where $r_{+}$ and $r_{-}$ are the outer and lower apparent horizons, respectively. In fact, the surfaces $r_{\pm}$ describe the exterior and interior apparent horizons of $\mathrm{V}^{5}$. The photons can escape from the outer apparent horizon and reach an arbitrary large distance, which will confirm that these surfaces are, in fact, apparent horizons, not the event horizons.

From Eqs.~(\ref{e(14a)})-(\ref{e(16a)}), the equations of motion for the angular coordinates $\theta$, $\phi$ and $\varphi$ have the following solutions: $\theta=\phi=\pi/2$ and $\dot{\varphi}=c_{4}/r^{2}$, where $c_{4}$ is a constant (of integration) describing (and playing the role of) the orbital angular momentum of the geodesic particles. By using these equations, we are going to produce different types of strong retractions in the five-dimensional Lovelock black hole spacetime, as follows.
%
%%%%%%%%%%%%%%%%%%%%%%%%%%%%%%%%%%%%%%%%%%%%%%%%%%%%%%%%%%%%%%%%%%%%%%%%%%%%%%%%%%%%%%%%%%%%%% Strong retractions
%
\subsection{Strong retractions}
Now, we will discuss the strong retractions, which describe the geodesic strong retractions of the $\mathrm{V}^{\mathrm{5}}$ apparent horizons. In order to apply Eq.~(\ref{e(16a)}), we will examine three different scenarios, namely, $\dot{\varphi}=0$, $\sin\theta=0$ and $\sin\phi=0$.

First, let us consider the case when $\dot{\varphi}=0$, as well as the case where $\varphi=0$. In this case, the parameters (\ref{e(8)}) are given by
\begin{eqnarray}
\underset{(\varphi=0)}{u_{1}} & = & \pm \sqrt{f(r)}t+b_{1},\nonumber\\
\underset{(\varphi=0)}{u_{2}} & = & \pm \biggl\{\frac{r}{\sqrt{3}}\sqrt{\frac{2a-6M+3r}{r}}+\biggl(M-\frac{a}{3}\biggr)\log\biggl(r\sqrt{\frac{6a-18M+9r}{r}}+3\biggr)+a-3M\biggr\}+b_{2},\nonumber\\
\underset{(\varphi=0)}{u_{3}} & = & \pm r\phi+b_{3},\nonumber\\
\underset{(\varphi=0)}{u_{4}} & = & \pm (r\sin\phi)\theta+b_{4},\nonumber\\
\underset{(\varphi=0)}{u_{5}} & = & b_{5}.
\label{e(15)}
\end{eqnarray}
These components represent the subspace geodesic $C_{(\varphi=0)}$ in $\mathrm{V}^{5}$, and therefore we get the main type of strong retraction $\varpi_{(\varphi =0)}:\mathrm{V}^{5} \rightarrow C_{(\varphi =0)}$.

On the other hand, in the case where $\sin\theta=0$, we obtain another subspace $C_{(\theta=0)}$ in $\mathrm{V}^{5}$, whose components are given by
\begin{eqnarray}
\underset{(\theta=0)}{u_{1}} & = & \pm \sqrt{f(r)}t+b_{1},\nonumber\\
\underset{(\theta=0)}{u_{2}} & = & \pm \biggl\{\frac{r}{\sqrt{3}}\sqrt{\frac{2a-6M+3r}{r}}+\biggl(M-\frac{a}{3}\biggr)\log\biggl(r\sqrt{\frac{6a-18M+9r}{r}}+3\biggr)+a-3M\biggr\}+b_{2},\nonumber\\
\underset{(\theta=0)}{u_{3}} & = & \pm r\phi+b_{3},\nonumber\\
\underset{(\theta=0)}{u_{4}} & = & b_{4},\nonumber\\
\underset{(\theta=0)}{u_{5}} & = & b_{5},
\label{e(16)}
\end{eqnarray}
which is another type of strong retraction, namely, $\varpi_{(\theta=0)}:\mathrm{V}^{5} \rightarrow C_{(\theta=0)}$.

Finally, we can compute the case when $\sin\phi=0$ and hence deduce the subspace $C_{(\phi=0)}$ in $\mathrm{V}^{5}$, whose components are given by
\begin{eqnarray}
\underset{(\phi=0)}{u_{1}} & = & \pm \sqrt{f(r)}t+b_{1},\nonumber\\
\underset{(\phi=0)}{u_{2}} & = & \pm \biggl\{\frac{r}{\sqrt{3}}\sqrt{\frac{2a-6M+3r}{r}}+\biggl(M-\frac{a}{3}\biggr)\log\biggl(r\sqrt{\frac{6a-18M+9r}{r}}+3\biggr)+a-3M\biggr\}+b_{2},\nonumber\\
\underset{(\phi=0)}{u_{3}} & = & b_{3},\nonumber\\
\underset{(\phi=0)}{u_{4}} & = & b_{4},\nonumber\\
\underset{(\phi=0)}{u_{5}} & = & b_{5},
\label{e(17)}
\end{eqnarray}
where the new type of strong retraction is therefore $\varpi_{(\phi=0)}:\mathrm{V}^{5} \rightarrow C_{(\phi=0)}$. Moreover, these types of strong retractions describe the apparent horizons of the five-dimensional Lovelock black hole spacetime $\mathrm{V}^{\mathrm{5}}$.
%
%%%%%%%%%%%%%%%%%%%%%%%%%%%%%%%%%%%%%%%%%%%%%%%%%%%%%%%%%%%%%%%%%%%%%%%%%%%%%%%%%%%%%%%%%%%%%% Strong homotopy retracts
%
\subsection{Strong homotopy retracts}
In what follows, we will use different types of strong retracts, such the ones mentioned above, to introduce the strong homotopy retract of the five-dimensional Lovelock black hole $\mathrm{V}^{5}$.

The strong homotopy retracts on $\mathrm{V}^{5}$ are defined as $\mathfrak{H}:\mathrm{V}^{5} \times [0,1] \rightarrow \mathrm{V}^{5}$. Thus, from the retractions, we obtain
\begin{eqnarray}
\varpi_{(\varphi=0)} & : & \mathrm{V}^{5} \rightarrow C_{(\varphi=0)},\nonumber\\
\varpi_{(\theta=0)} & : & \mathrm{V}^{5} \rightarrow C_{(\theta=0)},\nonumber\\
\varpi_{(\phi=0)} & : & \mathrm{V}^{5} \rightarrow C_{(\phi=0)},
\label{e(21)}
\end{eqnarray}
so that the strong homotopy retracts of $\mathrm{V}^{5}$ into a geodesics $C_{(\varphi=0)} \subseteq \mathrm{V}^{5}$ are given by
\begin{equation}
\mathfrak{H}_{C_{(\varphi=0)}}:\mathrm{V}^{5} \times [0,1] \rightarrow \mathrm{V}^{5},
\label{e(22)}
\end{equation}
where
\begin{equation}
\mathfrak{H}_{C_{(\varphi=0)}} (u,t)=(1+e^{-t})\frac{1}{2}\cos\biggl(\frac{\pi t}{2}\biggr)\mathfrak{H}_{C_{(\varphi=0)}} (u,0)+(1+e^{-t})\frac{1}{2}\sin\biggl(\frac{\pi t}{2}\biggr)\mathfrak{H}_{C_{(\varphi=0)}} (u,1),
\label{e(23)}
\end{equation}
which is valid $\forall\ u \in \mathrm{V}^{5}$ and $\forall\ t \in [0,1]$, with
\begin{eqnarray}
\mathfrak{H}_{C_{(\varphi=0)}} (u,0) & = & \{u_{1},u_{2},u_{3},u_{4},u_{5}\},\nonumber\\
\mathfrak{H}_{C_{(\varphi=0)}} (u,1) & = & \biggl\{\underset{(\varphi=0)}{u_{1}},\underset{(\varphi=0)}{u_{2}},\underset{(\varphi=0)}{u_{3}},\underset{(\varphi=0)}{u_{4}},\underset{(\varphi=0)}{u_{5}}\biggr\}.
\label{e(24)}
\end{eqnarray}

Furthermore, for the case of strong homotopy retract of $\mathrm{V}^{5}$ into a geodesic $C_{(\theta=0)} \subseteq \mathrm{V}^{5}$, we get
\begin{equation}
\mathfrak{H}_{C_{(\theta=0)}}:\mathrm{V}^{5} \times [0,1] \rightarrow \mathrm{V}^{5},
\label{e(25)}
\end{equation}
where
\begin{equation}
\mathfrak{H}_{C_{(\theta=0)}} (u,t)=(1+e^{-t})\frac{1}{2}\cos\biggl(\frac{\pi t}{2}\biggr)\mathfrak{H}_{C_{(\theta=0)}} (u,0)+(1+e^{-t})\frac{1}{2}\sin\biggl(\frac{\pi t}{2}\biggr)\mathfrak{H}_{C_{(\theta=0)}} (u,1),
\label{e(26)}
\end{equation}
which is valid $\forall\ u \in \mathrm{V}^{5}$ and $\forall\ t \in [0,1]$, with
\begin{eqnarray}
\mathfrak{H}_{C_{(\theta=0)}} (u,0) & = & \{u_{1},u_{2},u_{3},u_{4},u_{5}\},\nonumber\\
\mathfrak{H}_{C_{(\theta=0)}} (u,1) & = & \biggl\{\underset{(\theta=0)}{u_{1}},\underset{(\theta=0)}{u_{2}},\underset{(\theta=0)}{u_{3}},\underset{(\theta=0)}{u_{4}},\underset{(\theta=0)}{u_{5}}\biggr\}.
\label{e(27)}
\end{eqnarray}

Finally, we can obtain the strong homotopy retract of $\mathrm{V}^{5}$ into a geodesic $C_{(\phi=0)} \subseteq \mathrm{V}^{5}$. It is given by
\begin{equation}
\mathfrak{H}_{C_{(\phi=0)}}:\mathrm{V}^{5} \times [0,1] \rightarrow \mathrm{V}^{5},
\label{e(28)}
\end{equation}
where
\begin{equation}
\mathfrak{H}_{C_{(\phi=0)}} (u,t)=(1+e^{-t})\frac{1}{2}\cos\biggl(\frac{\pi t}{2}\biggr)\mathfrak{H}_{C_{(\phi=0)}} (u,0)+(1+e^{-t})\frac{1}{2}\sin\biggl(\frac{\pi t}{2}\biggr)\mathfrak{H}_{C_{(\phi=0)}} (u,1),
\label{e(29)}
\end{equation}
which is valid $\forall\ u \in \mathrm{V}^{5}$ and $\forall\ t \in [0,1]$, with
\begin{eqnarray}
\mathfrak{H}_{C_{(\phi=0)}} (u,0) & = & \{u_{1},u_{2},u_{3},u_{4},u_{5}\},\nonumber\\
\mathfrak{H}_{C_{(\phi=0)}} (u,1) & = & \biggl\{\underset{(\phi=0)}{u_{1}},\underset{(\phi=0)}{u_{2}},\underset{(\phi=0)}{u_{3}},\underset{(\phi=0)}{u_{4}},\underset{(\phi=0)}{u_{5}}\biggr\}.
\label{e(30)}
\end{eqnarray}
In these results, the components $u_{i}$ (with $i=1,\ldots,5$) are given by Eqs.~(\ref{e(8)}), (\ref{e(15)}), (\ref{e(16)}) and (\ref{e(17)}).%and $\underset{(\varphi,\theta,\phi=0)}{u_{i}}$ 

Therefore, from these results, we can conclude that all types of strong retractions on $\mathrm{V}^{5}$ induce strong homotopy retracts on the five-dimensional Lovelock black hole spacetime. In addition, we also conclude that all geodesic strong retractions on a cross-section of the five-dimensional Lovelock black hole apparent horizon are strongly homotopical retracted as subspaces and contained in $\mathrm{V}^{5}$. Thus, we have shown (and proved, from a mathematical point of view,) the existence of apparent horizons in the five-dimensional Lovelock black hole.

In what follows, we will calculate the perihelion precession and analyze the motion of quantum scalar particles outside of the exterior apparent horizon, as well as far from the black hole at the asymptotic infinity, which led to some interesting physics.
%
%%%%%%%%%%%%%%%%%%%%%%%%%%%%%%%%%%%%%%%%%%%%%%%%%%%%%%%%%%%%%%%%%%%%%%%%%%%%%%%%%%%%%%%%%%%%%% Perihelion precession
%
\section{Perihelion precession}\label{Perihelion}
Now, we are interested in the trajectory of massive particles in the equatorial plane. In this scenario, we will take the following choice to exploiting the spherical symmetry such that $\phi=\pi/2$, $\theta=\pi/2$, and $\varphi$ as the independent coordinate. Then, the (geodesic) equation of motion, after some algebraic manipulation, can be written as
\begin{equation}
\biggl(\frac{du}{d\varphi}\biggr)^{2}=\frac{1}{L^{2}}[E^{2}-f(u)(1+L^{2}u^{2})],
\label{eq:equation_geodesic}
\end{equation}
with
\begin{equation}
f(u)=1-Mu^{2}-\frac{2a}{3}u,
\label{eq:f_u}
\end{equation}
where we have defined $r=1/u$. Here, $E$ is the energy per particle mass, and $L$ is the angular momentum per particle mass. On further differentiation with respect to the angular coordinate $\varphi$ and keeping terms $\sim u^{2}$, we get
\begin{equation}
\frac{d^{2}u}{d\varphi^{2}}+(1+\epsilon_{1})u=A+\epsilon_{2}u^{2},
\label{eq:differentiation}
\end{equation}
where $\epsilon_{1}=-M/L^{2}$, $\epsilon_{2}=a$, and $A=a/3L^{2}$. Then, we assume a periodic solution, that is, $\varphi_{\rm periodic}=2\pi(1+\epsilon_{2}\alpha)$, where $\alpha$ is the measure of the perihelion precession. By using a perturbation method, the solution for $u$ can be taken as $u=C+D\cos(1-\epsilon_{2}\alpha)\varphi+\epsilon_{2} u_{1}$, where $u_{1}=u_{1}(\varphi)$. Therefore, by keeping the terms linear in $\epsilon_{2}$, we obtain the following expression for the precession angle:
\begin{equation}
\delta\varphi=2 \pi A \epsilon_{2}-\pi\epsilon_{1}=\frac{\pi}{L^{2}}\biggl(M+\frac{2a^{2}}{3}\biggr).
\label{eq:precession_angle}
\end{equation}
From Eq.~(\ref{eq:precession_angle}), we can see that the perihelion shift is directly proportional to the total mass $M$, as well as to the string cloud parameter $a$. It is worth emphasizing that this solution does not reduce to the Schwarzschild one in the limit when $a \rightarrow 0$, since the action in the five-dimensional Lovelock spacetime is not the usual Einstein-Hilbert action even if $a=0$, so that we can call this result as an anomalous perihelion precession. In addition, we can mention that this expression for the perihelion shift is very similar to the one concerning the Reissner-Nordstr\"{o}m black hole \cite{PhysRevD.89.026003} but with different signs, as well as its behavior is analog to the one concerning the Garfinkle-Horowitz-Strominger dilaton black hole \cite{PhysRevD.89.026003} but with different amplitudes.

\begin{figure}%[ht]
\centering
\includegraphics[width=1\columnwidth]{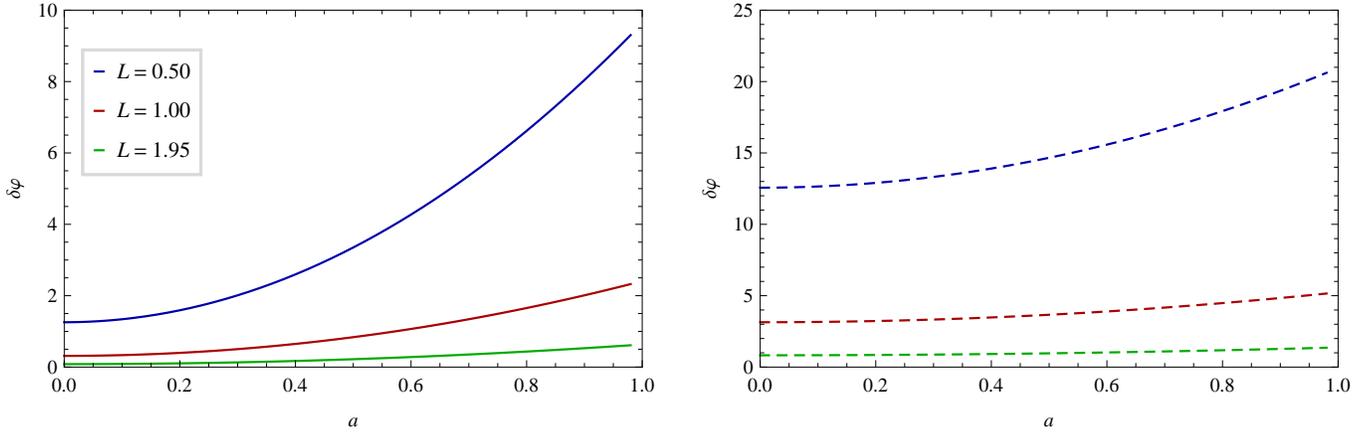}
\caption{The variation of the perihelion shift. We focus on the cases $M=0.1$ (solid lines), and $M=1.0$ (dashed lines).}
\label{fig:Homotopy_5D_Lovelock_Fig1}
\end{figure}

The behavior of the perihelion precession $\delta\varphi$ is shown in Fig.~\ref{fig:Homotopy_5D_Lovelock_Fig1}, for different values of the total mass, as functions of the string cloud parameter $a$.

%
%%%%%%%%%%%%%%%%%%%%%%%%%%%%%%%%%%%%%%%%%%%%%%%%%%%%%%%%%%%%%%%%%%%%%%%%%%%%%%%%%%%%%%%%%%%%%% Quasistationary levels and wave functions
%
\section{Quasistationary levels and wave functions}\label{Quasistationary}
In this section, we will discuss the motion of quantum scalar particles in the five-dimensional Lovelock black hole spacetime. To do this, we will generalize the previous results obtained by Vieira \cite{Vieira}, in the sense that now we will obtain an analytical solution for the radial part of the Klein-Gordon equation that can be used to study the quasistationary levels in the background under consideration.

The quasistationary levels, also known as quasibound states or resonance spectra, are obtained by imposing two boundary conditions on the radial solution, namely, it should describe an ingoing wave at the exterior apparent horizon and tend to zero far from the black hole at asymptotic infinity. The general method used to obtain these resonant frequencies was described by Vieira and Kokkotas \cite{PhysRevD.104.024035}. Here, we will just show our results and cordially invite the readers to find the method fully described in Ref.~\cite{PhysRevD.104.024035}.

\begin{figure}%[ht]
	\centering
		\includegraphics[scale=1.00]{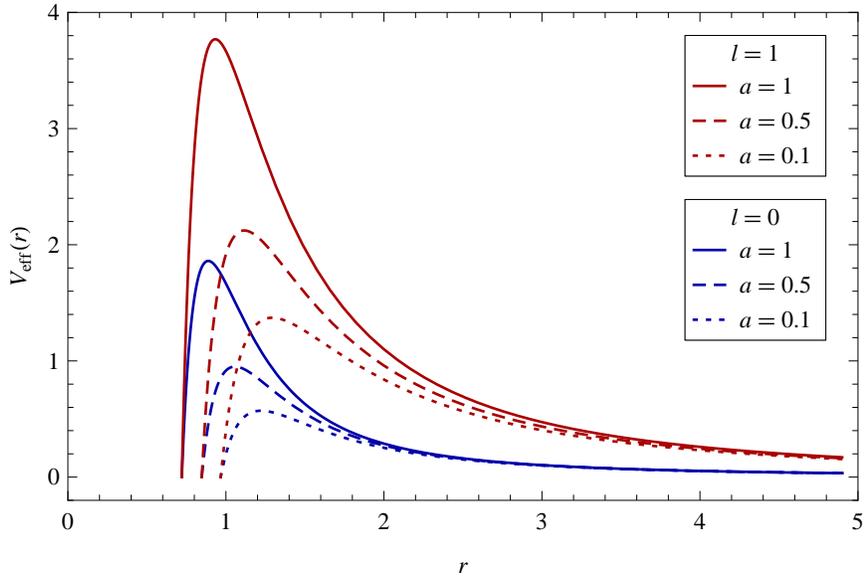}
	\caption{The effective potentials for $M=1$ and $\mu=0$, and different values of the azimuthal quantum number $l(=0,1)$ and the string cloud parameter $a(=0.1,0.5,1)$.}
	\label{fig:Homotopy_5D_Lovelock_Fig2}
\end{figure}

Thus, let us write the Klein-Gordon equation as
\begin{equation}
\biggl[\frac{1}{\sqrt{-g}}\partial_{\sigma}(g^{\sigma\tau}\sqrt{-g}\partial_{\tau})-\mu^{2}\biggr]\Psi=0,
\label{eq:Klein-Gordon}
\end{equation}
where $\mu$ is the mass of the scalar particle. The wave function $\Psi$ can be separated as $\Psi=U(r)Y_{slm}(\phi,\theta,\varphi)\mbox{e}^{-i \omega t}$, where $\omega$ is the frequency of the scalar particle, and $Y_{slm}(\phi,\theta,\varphi)$ is the three-dimensional normalized spherical harmonic function, in which $m$ is the azimuthal quantum number, $l$ is the orbital quantum number and $s$ is the angular quantum number related to the five dimension. Here, $U(r)=R(r)/r^{3/2}$ is the radial function, which satisfy the following equation
\begin{equation}
\frac{d^{2} R(r)}{d r^{2}}+\frac{1}{f(r)}\frac{d f(r)}{d r}\frac{d R(r)}{d r}+\biggl\{\frac{\omega^{2}}{[f(r)]^{2}}-\frac{1}{4r^{2}f(r)}\biggl[4\lambda_{slm}+4r^{2}\mu^{2}+3f(r)+6r\frac{d f(r)}{d r}\biggr]\biggr\}R(r)=0,
\label{eq:radial_5D_Lovelock}
\end{equation}
where $\lambda_{slm}=s(s+2)$ is the separation constant (an angular eigenvalue). Now, by introducing the ``tortoise coordinate'' $r_{*}$, which is defined by the relation $dr_{*}=dr/f(r)$, we can write this radial function as
\begin{equation}
\frac{d^{2} R(r)}{d r_{*}^{2}}+[\omega^{2}-V_{\rm eff}(r)]R(r)=0,
\label{eq:radial_tortoise_5D_Lovelock}
\end{equation}
where $V_{\rm eff}(r)$ is the effective potential and given by
\begin{equation}
V_{\rm eff}(r)=f(r)\biggl[\frac{\lambda_{slm}}{r^{2}}+\mu^{2}+\frac{3f(r)}{4r^{2}}+\frac{3}{2r}\frac{df(r)}{dr}\biggr].
\label{eq:Veff_5D_Lovelock}
\end{equation}
The behavior of the effective potential $V_{\rm eff}(r)$ is shown in Fig.~\ref{fig:Homotopy_5D_Lovelock_Fig2}, for different values of the azimuthal quantum number and the string cloud parameter.

From Eq.~(\ref{eq:Veff_5D_Lovelock}) and Fig.~\ref{fig:Homotopy_5D_Lovelock_Fig2}, we conclude that the width of the potential barrier increases with both the azimuthal quantum number $l$ and the string cloud parameter $a$. For the chosen values, the exterior apparent horizon $r_{+}$ is located at $r=0.720759$, $r=0.847127$ and $r=0.967222$ for $a=1$, $a=0.5$ and $a=0.1$, respectively.

Then, we can provide an analytical solution for the radial equation (\ref{eq:radial_5D_Lovelock}) in terms of the general Heun functions \cite{Ronveaux:1995}. It is given by
\begin{eqnarray}
R(x) & = & x^{\frac{\gamma-1}{2}}(x-1)^{\frac{\delta-1}{2}}(x-b)^{\frac{\epsilon}{2}}\nonumber\\
		 &   & \times \{C_{1}\ \mbox{HeunG}(b,q;\alpha,\beta,\gamma,\delta;x) + C_{2}\ x^{1-\gamma}\ \mbox{HeunG}(b,q_{2};\alpha_{2},\beta_{2},\gamma_{2},\delta;x)\},
\label{eq:general_solution_radial_x_5D_Lovelock}
\end{eqnarray}
with
\begin{equation}
x=\frac{r-r_{+}}{r_{-}-r_{+}}
\label{eq:radial_coordinate_x_5D_Lovelock}
\end{equation}
and
\begin{equation}
b=\frac{r_{+}}{r_{+}-r_{-}},
\label{eq:singularity_b_5D_Lovelock}
\end{equation}
where $C_{1}$ and $C_{2}$ are constants to be determined, and $\mbox{HeunG}(b,q;\alpha,\beta,\gamma,\delta;x)$ is the general Heun function. The parameters $\alpha$, $\beta$, $\gamma$, $\delta$, $\epsilon$ and $q$ are given by
\begin{equation}
\alpha=-\sqrt{1-\frac{\lambda_{slm}}{M}}+\sqrt{5+\mu^{2}+2\sqrt{1-\frac{\lambda_{slm}}{M}}}-3i\omega\sqrt{\frac{1}{a^{2}+9M}},
\label{eq:alpha_radial_x_5D_Lovelock}
\end{equation}
\begin{equation}
\beta=-\sqrt{1-\frac{\lambda_{slm}}{M}}-\sqrt{5+\mu^{2}+2\sqrt{1-\frac{\lambda_{slm}}{M}}}-3i\omega\sqrt{\frac{1}{a^{2}+9M}},
\label{eq:beta_radial_x_5D_Lovelock}
\end{equation}
\begin{equation}
\gamma=1-3i\omega\sqrt{\frac{1}{a^{2}+9M}},
\label{eq:gamma_radial_x_5D_Lovelock}
\end{equation}
\begin{equation}
\delta=1-3i\omega\sqrt{\frac{1}{a^{2}+9M}},
\label{eq:delta_radial_x_5D_Lovelock}
\end{equation}
\begin{equation}
\epsilon=-1-2\sqrt{1-\frac{\lambda_{slm}}{M}},
\label{eq:eta_radial_x_5D_Lovelock}
\end{equation}
\begin{equation}
q=\frac{(r_{+}-r_{-})^2 \{r_{+}^2[\gamma(\delta+\epsilon)-2(\mu^2+2)]-2\lambda_{slm}+r_{+}r_{-}(3-\gamma\epsilon)\}-4r_{+}^2\omega^2}{2r_{+}(r_{+}-r_{-})^3}.
\label{eq:q_radial_x_5D_Lovelock}
\end{equation}
Finally, the parameters $\alpha_{2}$, $\beta_{2}$, $\gamma_{2}$ and $q_{2}$, related to the second solution, are defined as
\begin{equation}
\alpha_{2}=\alpha+1-\gamma,
\label{eq:alpha_2_general_Heun}
\end{equation}
\begin{equation}
\beta_{2}=\beta+1-\gamma,
\label{eq:beta_2_general_Heun}
\end{equation}
\begin{equation}
\gamma_{2}=2-\gamma.
\label{eq:gamma_2_general_Heun}
\end{equation}
\begin{equation}
q_{2}=q+(\alpha\delta+\epsilon)(1-\gamma).
\label{eq:q_2_general_Heun}
\end{equation}

Now, the first boundary condition for quasibound states implies that we should keep only the first (linearly independent) radial solution, i.e., we have to choose $C_{2}=0$ in order to get
\begin{equation}
\lim_{r \rightarrow r_{+}} R(r) \sim C_{1}\ R^{\rm in},
\label{eq:radial_horizon_5D_Lovelock}
\end{equation}
where $R^{\rm in}=(r-r_{+})^{-\frac{i\omega}{2\kappa_{+}}}$, with $\kappa_{+}(=\sqrt{a^{2}+9M}/3)$ being the gravitational acceleration on the exterior apparent horizon $r_{+}$. Next, the second boundary condition for quasibound states implies that the radial wave solution behaves as
\begin{equation}
\lim_{r \rightarrow \infty} R(r) \sim C_{1}\ r^{\sigma},
\label{eq:radial_infinity_5D_Lovelock}
\end{equation}
with $\sigma=D-\alpha$, where the coefficient $D$ is given by
\begin{equation}
D=-\frac{1}{2}-\sqrt{1-\frac{s(s+2)}{M}}-3i\omega\sqrt{\frac{1}{a^{2}+9M}}.
\label{eq:D_radial_infinity_5D_Lovelock}
\end{equation}
The behavior of the radial wave solution, far from the black hole at asymptotic infinity, is determined by analyzing the sign of the real part of $\sigma$: $\mbox{Re}[\sigma] > 0$ implies that the radial function diverges, while $\mbox{Re}[\sigma] < 0$ implies that the radial function tends to zero describing the quasibound states. Thus, the final behavior of the radial wave solution will be determined according to the values of the resonant frequencies $\omega$, which will be obtained in what follows.

The scalar resonant frequencies describing quasibound states in the five-dimensional Lovelock black hole can be obtained by imposing the so-called polynomial alpha-condition related to the general Heun functions, which is given by
\begin{equation}
\alpha=-n,
\label{eq:alpha-condition}
\end{equation}
where $n=0,1,2,\ldots$ is now the principal quantum number. This condition gives the following set of resonant frequencies
\begin{equation}
\omega_{sn}=-i\frac{\sqrt{a^{2}+9M}}{3}\biggl[\sqrt{5+\mu^{2}+2\sqrt{1-\frac{s(s+2)}{M}}}-\sqrt{1-\frac{s(s+2)}{M}}+n\biggr].
\label{eq:omega_5D_Lovelock}
\end{equation}
In Table \ref{tab:I_5D_Lovelock} we present some values of the resonant frequencies $\omega_{sn}$, as well as the corresponding coefficients $\sigma_{sn}(=D_{sn}+n)$, as functions of the string cloud parameter $a$.

\newpage

\begin{table}%[ht]
\caption{The resonant frequencies $\omega_{sn}$ and the real part of the corresponding coefficients $\sigma_{sn}$ for $M=1$ and $\mu=0.1$. We focus on the mode $s=0$ for $n=0,1$.}
\label{tab:I_5D_Lovelock}
\begin{tabular}{c||c|c||c|c}
\hline\noalign{\smallskip}
			$a$   & $\omega_{00}$ & $\mbox{Re}[\sigma_{00}]$ & $\omega_{01}$ & $\mbox{Re}[\sigma_{01}]$ \\
\noalign{\smallskip}\hline\noalign{\smallskip}
			$0.0$ & $-1.647640i$  & $-3.147640$              & $-2.647640i$  & $-3.147640$              \\
			$0.1$ & $-1.648556i$  & $-3.147640$              & $-2.649111i$  & $-3.147640$              \\
			$0.2$ & $-1.651298i$  & $-3.147640$              & $-2.653518i$  & $-3.147640$              \\
			$0.3$ & $-1.655858i$  & $-3.147640$              & $-2.660846i$  & $-3.147640$              \\
			$0.4$ & $-1.662222i$  & $-3.147640$              & $-2.671071i$  & $-3.147640$              \\
			$0.5$ & $-1.670368i$  & $-3.147640$              & $-2.684161i$  & $-3.147640$              \\
			$0.6$ & $-1.680270i$  & $-3.147640$              & $-2.700074i$  & $-3.147640$              \\
			$0.7$ & $-1.691898i$  & $-3.147640$              & $-2.718760i$  & $-3.147640$              \\
			$0.8$ & $-1.705217i$  & $-3.147640$              & $-2.740162i$  & $-3.147640$              \\
			$0.9$ & $-1.720187i$  & $-3.147640$              & $-2.764218i$  & $-3.147640$              \\
			$1.0$ & $-1.736766i$  & $-3.147640$              & $-2.790858i$  & $-3.147640$              \\
\noalign{\smallskip}\hline
\end{tabular}
\end{table}

In Table \ref{tab:I_5D_Lovelock} we see that the resonant frequencies $\omega_{sn}$ are physically admissible (in the mode under consideration), which represent the quasibound states of scalar particles in the five-dimensional Lovelock black hole spacetime. In this scenario, the radial wave function, given by Eq.~(\ref{eq:general_solution_radial_x_5D_Lovelock}) with $C_{2}=0$, tends to zero far from the black hole at asymptotic infinity, since $\mbox{Re}[\sigma_{sn}] < 0$, which satisfy the required conditions for quasibound states. Furthermore, we can conclude that the motion is over-damped (with purely imaginary frequencies) in this mode ($s=0$). Therefore, this result may describe a stable system, since the resonant frequencies are always negative.

\begin{figure}%[ht]
\centering
\includegraphics[width=1\columnwidth]{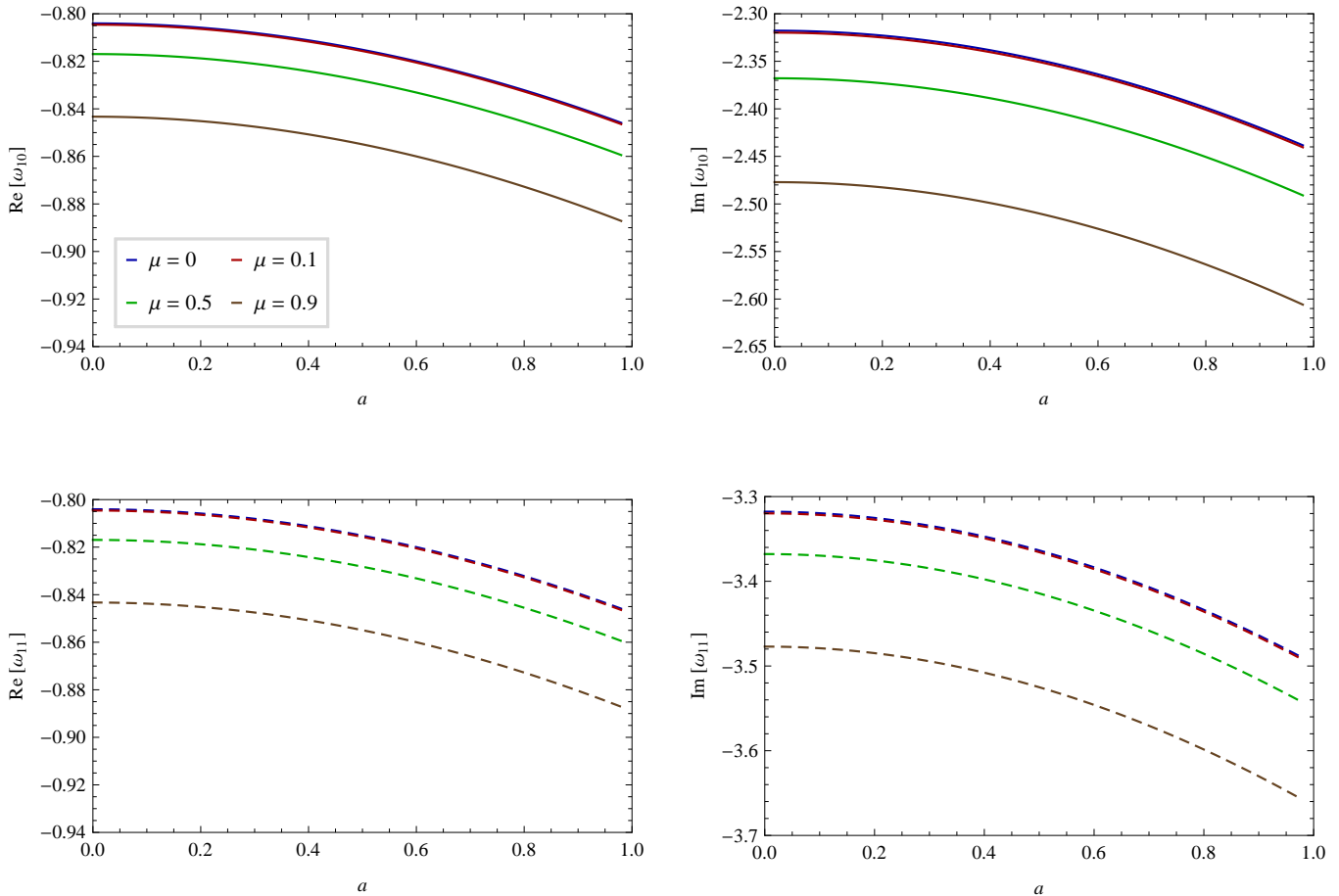}
\caption{The resonant frequencies in the five-dimensional Lovelock black hole spacetime. The left plots show the oscillation frequency ($\mbox{Re}[\omega_{sn}]$), while the right plots show the decay rate ($\mbox{Im}[\omega_{sn}]$). We focus on the mode $s=1$ for $n=0,1$ (solid and dashed lines, respectively), where $M=1$.}
\label{fig:Homotopy_5D_Lovelock_Fig3}
\end{figure}

The behavior of the massless resonant frequencies $\omega_{sn}$ is shown in Fig.~\ref{fig:Homotopy_5D_Lovelock_Fig3}, for different values of the mass of scalar particle, as functions of the string cloud parameter $a$.

From Fig.~\ref{fig:Homotopy_5D_Lovelock_Fig3} we conclude that both real and imaginary parts of the resonant frequencies $\omega_{sn}$ increase (in modulus) as the string cloud parameter $a$ grows (in the mode under consideration). In these modes, the motion is simply damped. In addition, we may conclude that the five-dimensional Lovelock black holes are stable, since the imaginary part of the resonant frequencies is always negative.

In addition, we can compute the transition frequency, $\Delta \omega$, between two highly damped ($n \rightarrow \infty$) neighboring states \cite{arXiv:2107.02065}. It is given by
\begin{equation}
\Delta \omega \approx \mbox{Im}[\omega_{s(n-1)}]-\mbox{Im}[\omega_{sn}] = \frac{\sqrt{a^{2}+9M}}{3}.
\label{eq:transition_5D_Lovelock}
\end{equation}
This means that $\Delta\omega=\kappa_{+}$ and hence the area of the black hole exterior apparent horizon is quantized, that is, the area spectrum is given by
\begin{equation}
\mathcal{A}_{+_n} = 8 \pi n \hbar.
\label{eq:area_spectrum_5D_Lovelock}
\end{equation}

\begin{figure}%[ht]
\centering
\includegraphics[width=1\columnwidth]{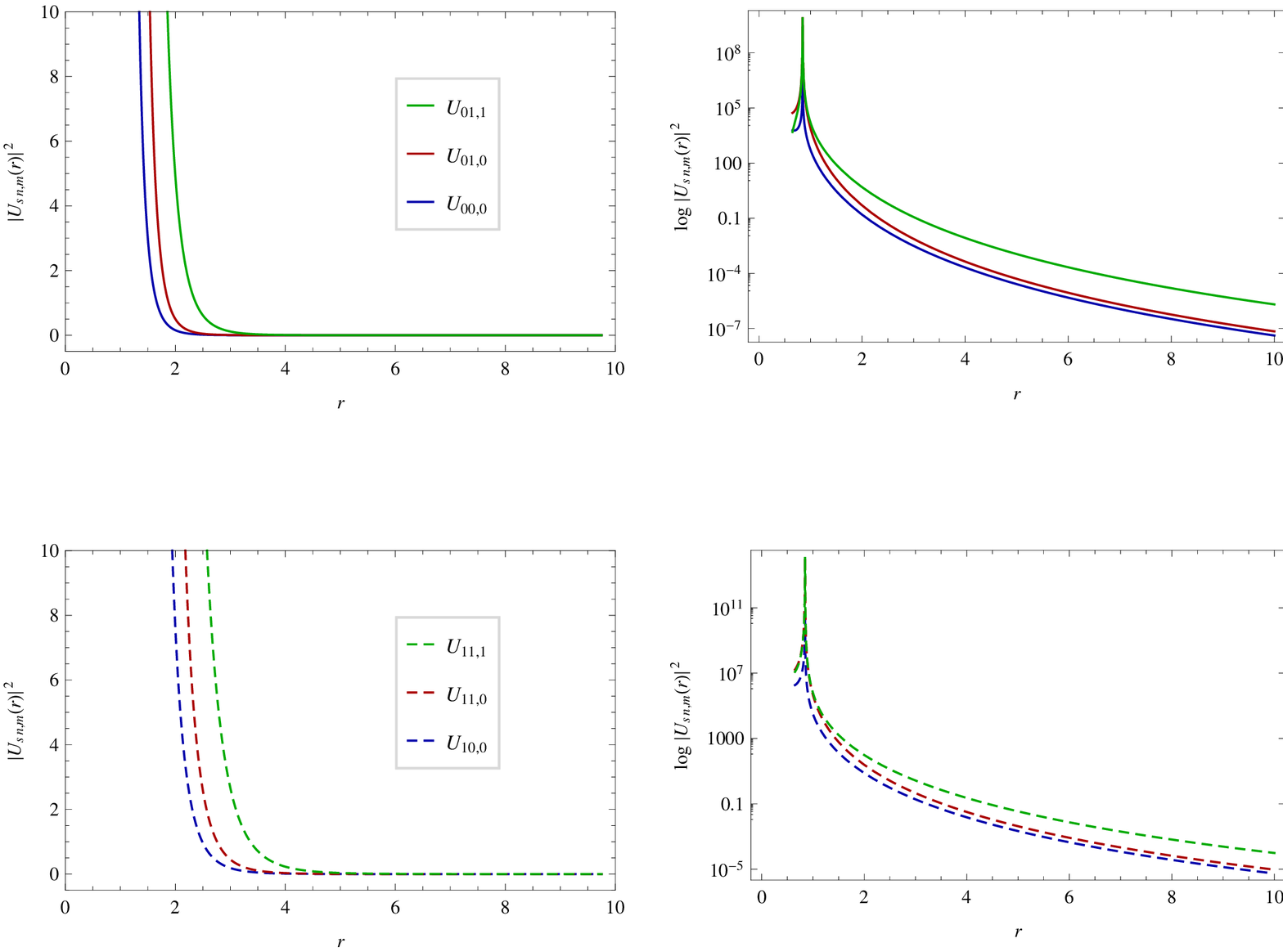}
\caption{The first three squared radial wave eigenfunctions $U_{sn,m}(r)=R_{sn,m}(r)/r^{3/2}$. The units are in multiples of $C_{sn,m}$. We focus on the modes $s=0,1$ (solid and dashed lines, respectively), where $M=1$, $\mu=0.1$ and $a=0.5$.}
\label{fig:Homotopy_5D_Lovelock_Fig4}
\end{figure}

Now, we be still using the Vieira-Kokkotas method \cite{PhysRevD.104.024035} to derive the radial wave eigenfunctions describing scalar particles that propagate in the five-dimensional Lovelock black hole spacetime. This method consists of using some properties of the general Heun functions, as well as their polynomial alpha-condition, to write the general Heun polynomials and then obtains the radial wave eigenfunctions. Once again, we cordially invite the readers to find the method fully described in Ref.~\cite{PhysRevD.104.024035}. Thus, the radial wave eigenfunctions can be written as
\begin{equation}
R_{sn,m}(x)=C_{sn,m}\ x^{\frac{\gamma-1}{2}}(x-1)^{\frac{\delta-1}{2}}(x-b)^{\frac{\epsilon}{2}}\ \mbox{Hp}_{sn,m}(x),
\label{eq:eigenfunctions_5D_Lovelock}
\end{equation}
where $C_{sn,m}$ is a constant to be determined, $\mbox{Hp}_{sn,m}(x)$ are the general Heun polynomials, and $m=0,\ldots,n$ is a parameter related to the order of the polynomials. The first three squared radial wave eigenfunctions are presented in Fig.~\ref{fig:Homotopy_5D_Lovelock_Fig4}.

From Fig.~\ref{fig:Homotopy_5D_Lovelock_Fig4} we can conclude that the resonant frequencies $\omega_{sn}$ are quasibound states, since their radial wave eigenfunctions tend to zero at infinity and diverge at the exterior apparent horizon; from the right plots, we see that it mathematically reaches a maximum value and then crosses into the black hole.
%
%%%%%%%%%%%%%%%%%%%%%%%%%%%%%%%%%%%%%%%%%%%%%%%%%%%%%%%%%%%%%%%%%%%%%%%%%%%%%%%%%%%%%%%%%%%%%% Conclusions
%
\section{Conclusions}\label{Conclusions}
This paper characterizes a modern application of mathematical concepts on geometric topology, and physics as well, specifically, on the strong homotopy retracts in the five-dimensional Lovelock black hole spacetime. From the metric describing the background under consideration, we have shown that the geodesics are strong retractions.

We investigated, through the theory of retracts, the role played, in principle, by the topology in the mathematical physics studies of spacetime singularities, in particular, the apparent horizons of the five-dimensional Lovelock black holes. It is worth emphasizing that our understanding about the spacetime singularities is still incomplete, since we need a completely consistent quantum theory of gravity to do this. We presented some results concerning a higher dimensional black hole metric, but these studies can be also applied for any gravitational system such as compact objects, as well as to the cosmological universe.

We calculated the perihelion precession and found that it describes a kind of anomalous effect (when compared with the known cases in the literature) in the motion of massive particles in the equatorial plane.

In this work, we also analyzed the behavior of quantum scalar fields propagating near the exterior apparent horizon of the five-dimensional Lovelock black hole spacetime. We obtained an analytical solution for the radial part of the Klein-Gordon equation and then examined a very important phenomenon related to the boundary conditions imposed to it, namely, the radial wave eigenfunctions.

In what concerns to the radial wave eigenfunctions, we got their well behaved form, which is given in terms of the general Heun polynomials. From these results, we conclude that there exist quasistationary levels (or quasibound states as described by Vieira and Kokkotas \cite{PhysRevD.104.024035}) in the background under consideration.
%
%%%%%%%%%%%%%%%%%%%%%%%%%%%%%%%%%%%%%%%%%%%%%%%%%%%%%%%%%%%%%%%%%%%%%%%%%%%%%%%%%%%%%%%%%%%%%% Author contributions
%
\section*{Author contributions}
M. Abu-Saleem: Conceptualization, Methodology, Investigation, Writing--original draft.

H. S. Vieira: Software, Data curation, Visualization, Supervision, Validation, Writing--review and editing.
%
%%%%%%%%%%%%%%%%%%%%%%%%%%%%%%%%%%%%%%%%%%%%%%%%%%%%%%%%%%%%%%%%%%%%%%%%%%%%%%%%%%%%%%%%%%%%%% Data availability
%
\section*{Data availability}
The data that support the findings of this study are available from the corresponding author upon reasonable request.
%
%%%%%%%%%%%%%%%%%%%%%%%%%%%%%%%%%%%%%%%%%%%%%%%%%%%%%%%%%%%%%%%%%%%%%%%%%%%%%%%%%%%%%%%%%%%%%% Acknowledgments
%
\begin{acknowledgments}
H.S.V. is funded by the Alexander von Humboldt-Stiftung/Foundation (Grant No. 1209836). This study was financed in part by the Coordena\c c\~{a}o de Aperfei\c coamento de Pessoal de N\'{i}vel Superior - Brasil (CAPES) - Finance Code 001.
\end{acknowledgments}
%
%%%%%%%%%%%%%%%%%%%%%%%%%%%%%%%%%%%%%%%%%%%%%%%%%%%%%%%%%%%%%%%%%%%%%%%%%%%%%%%%%%%%%%%%%%%%%% thebibliography
%

%
%%%%%%%%%%%%%%%%%%%%%%%%%%%%%%%%%%%%%%%%%%%%%%%%%%%%%%%%%%%%%%%%%%%%%%%%%%%%%%%%%%%%%%%%%%%%%%
%
\end{document}